\documentclass[a4paper,12pt]{article}
\usepackage[pctex32]{graphics}

\newcommand{\be}{\begin{equation}}
\newcommand{\ee}{\end{equation}}
\newcommand{\ben}{\begin{eqnarray}\displaystyle}
\newcommand{\een}{\end{eqnarray}}

\begin{titlepage}

\begin{document}
{\baselineskip20pt


\vskip .6cm

\begin{center}
{\Large \bf Thermodynamics of Gauss-Bonnet black holes revisited}

\end{center} }

\vskip .6cm
 \centerline{\large Yun Soo Myung$^{1,a}$,
 Yong-Wan Kim $^{1,b}$,
and Young-Jai Park$^{2,c}$}

\vskip .6cm

\begin{center}
{$^{1}$Institute of Basic Science and School of Computer Aided
Science,
\\Inje University, Gimhae 621-749, Korea \\}

{$^{2}$Department of Physics, Sogang University, Seoul 121-742, Korea}
\end{center}

\vspace{5mm}


\begin{abstract}
We investigate the Gauss-Bonnet black hole in five dimensional
anti-de Sitter spacetimes (GBAdS). We analyze all thermodynamic
quantities of the GBAdS, which is characterized by the
Gauss-Bonnet coupling $c$ and mass $M$, comparing with those of
the Born-Infeld-AdS (BIAdS), Reissner-Norstr\"om-AdS black holes
(RNAdS), Schwarzschild-AdS (SAdS), and BTZ black holes. For $c<0$
we cannot obtain the black hole with positively definite
thermodynamic quantities of mass, temperature, and entropy because
the entropy does not satisfy the area-law.
On the other hand, for $c>0$, we find  the BIAdS-like black hole,
showing that the coupling $c$ plays the role of pseudo-charge.
Importantly, we could not obtain the SAdS in the limits of $c\to 0$,
which means  that the GBAdS is basically different from the SAdS.
In addition, we clarify the connections between thermodynamic
and dynamical stability.
Finally, we also conjecture that if a black hole is big and
thus globally stable, its quasinormal modes may take analytic expressions.

\end{abstract}


\vskip 0.8cm

\noindent $^a$ysmyung@inje.ac.kr \\
\noindent $^b$ywkim65@gmail.com \\
\noindent $^c$yjpark@sogang.ac.kr

\noindent
\end{titlepage}

\setcounter{page}{2}

\section{Introduction}
Black holes are very important objects in classical and quantum
gravity. One of the most important and profound properties in
black hole physics is its thermodynamics.
Since the black hole thermodynamics has a deep connection with
quantum mechanics of gravity, we have a natural question whether
the properties related with the black hole thermodynamics are
modified, if we add higher dimensional corrections to
Einstein-Hilbert action, which are expected to appear in an
effective theory of quantum gravity, for example, string theory.

The Gauss-Bonnet (GB) term is the lowest dimensional term
among higher dimensional correction ones. The spherically symmetric black hole
solutions are known~\cite{BD}, and the thermodynamics is also
calculated~\cite{MS} in Einstein-Gauss-Bonnet (EGB) theory. In
particular, in the nontrivial lowest five dimensions, the most
general theory of gravity leading to second order field equations
for the metric is the  EGB theory with the coupling constant $c$.
The GB term with $c=1/2\pi \alpha'>0$ appears as the
first curvature stringy correction to general
relativity~\cite{GW,Myers:1987yn}, when assuming that the tension
$\alpha'$ of a string is large as compared to the energy scale of
other variables. Recently, the study of black holes with higher derivative
curvature in anti-de Sitter (AdS) spaces has been considered by
many authors. Static AdS black hole solutions
in EGB gravity denote GBAdS in this work, presenting a number of
interesting features (see $e.g.$ \cite{wheeler}, \cite{Cho},
\cite{TM} and the references therein).

Since the pioneering work of Hawking-Page phase transition (HP2)
between thermal AdS and SAdS in four dimensions~\cite{HP}, the
research of the black hole thermodynamics has recently improved.
Moreover, the HP2 in five dimensions was discussed for string
theories~\cite{Witt}. In the HP2, one generally starts with
thermal radiation in AdS space appearing a small black hole with
negative heat capacity (SBH$_-$). Then, since the heat capacity
changes from negative infinity to positive infinity at the minimum
temperature, the large black hole with positive heat capacity
(LBH$_+$) finally comes out as a stable object. Evidently, there
is a change of the dominance at the critical temperature: from
thermal radiation  to black hole~\cite{HP}.  On the other hand, it
was  suggested that there exists a different phase transition
(HP1) between the small black hole with positive heat capacity
(SBH$_+$) and LBH$_+$ in the RNAdS for fixed charge
$Q<Q_c$~\cite{CEJM,DMMS1,Myu} and GBAdS~\cite{Cho,DMMS2}.

Recently, we have  obtained all thermodynamic quantities of the 3D
Einstein-Born-Infeld black holes, which are nonlinear
generalization of BTZ black holes~\cite{mkp3}. Furthermore, we
have proposed that the 4D Born-Infeld-anti-de Sitter black holes
(BIAdS) with the coupling constant $b$  have the main feature of
RNAdS with charge $Q$: two horizons and a degenerate horizon
because the BIAdS is a nonlinear generalization of the RNAdS  and
disconnects to the SAdS~\cite{mkp4}. These studies may enhance the
level of understanding the GBAdS.

On the other hand, some issues of 5D Gauss-Bonnet black holes
remain unclarified. These include thermodynamic stability,
dynamical stability, and the existence of quasinormal (QN) 
modes~\cite{Kokk,BSS,BM,HH,Siop,ML,MSIO,BC,GB}. Importantly, their
connection known as the correlated stability conjecture was not
clearly understood until now  because the sign of coupling $c$ was
not taken seriously for these analysis. String theories always
predict positive coupling of $c>0$, while the negative coupling of
$c<0$ is also available to study their black holes. At the first
sight, the former may give rise to a black hole like the RNAdS
with the charge $c$, while the latter may provide the SAdS with
negative mass $c$. However, this interpretation of $c$ is not
correct. For $c>0$ case,  the $bQ=0.5$ BIAdS will be used  to
study the $0<c<l^2/36$ GBAdS because their thermodynamic
properties are the nearly same. In this case, $c$ plays the role
of a pseudo-charge. The role of charge ``$c$" becomes clear when
considering the GBRNAdS black hole.  For $c<0$, the role of negative mass
``$c$" becomes clear when introducing the $k=-1$ topological GBAdS
(TGBAdS).

 In this paper, we address these issues for the GBAdS in five dimensions.
We revisit all thermodynamic quantities of the GBAdS,
which is characterized by the GB coupling $c$ and mass $M$,
comparing with those of the BIAdS, RNAdS, SAdS, and BTZ black holes.
The similar and related works on the thermodynamics of the GBAdS
were carried out in Ref. \cite{DMMS2,Hira}.
We point out that the thermodynamic properties
of GBAdS are closely related to their dynamical stabilities and
the expressions of QN frequencies.

The organization of this work is as follows. In Sec. 2, we analyze
the possible GBAdS black hole solutions depending on the GB coupling $c$.
In Sec. 3, we revisit all thermodynamic properties of the GBAdS  by
comparing those of the BIAdS, RNAdS, SAdS, and BTZ black holes. We
also analyze the thermodynamic stability comparing with the
approaches of the dynamic stability and quasi-normal modes in the
GBAdS black holes in Sec. 4 and Sec. 5, respectively. Finally, we
summarize and discuss our results in Sec. 6. In Appendix A, we
comment on thermodynamic properties of the $k=-1$ TGBAdS. In
Appendix B, we mention thermodynamic properties of the GBRNAdS
black hole.

\section{Structure of TGBAdS black holes}

Now, let us consider a 5D gravitational action in the presence of
a negative cosmological constant $\Lambda = -6/l^2$ and GB term as
\begin{equation}\label{baction}
I = \frac{1}{16\pi G_5} \int \! d^5 x \sqrt{- g}\left[ R -2\Lambda +\frac{c}{2} L_{GB} \right],
\end{equation}
where
\begin{equation}
L_{GB} =  R^2 -4 R_{\mu\nu} R^{\mu\nu} + R_{\mu\nu\rho\sigma} R^{\mu\nu\rho\sigma}.
\label{GBaction}
\end{equation}
Here $G_5$ is the Newton constant and $c$ is a GB coupling
constant having mass dimension $-2$. In this work, we consider
both the cases with $c>0$ and $c<0$ comparing with the $c=0$ case
of SAdS. This action possesses black hole solutions, which we call
GBAdS \cite{BD, Cho, TM, RM, NO, RCAI, CNO}.

Varying the action (\ref{baction}) with respect to the
gravitational field $g_{\mu\nu}$, the field equation is obtained
as follows
\begin{equation}
\label{eqs} R_{\mu \nu } -\frac{1}{2}Rg_{\mu \nu}+\Lambda g_{\mu
\nu }+\frac{c}{2}H_{\mu \nu}=0,
\end{equation}
where
\begin{equation}
\label{eq1}
H_{\mu \nu}=2(R_{\mu \sigma \kappa \tau }R_{\nu }^{\phantom{\nu}%
\sigma \kappa \tau }-2R_{\mu \rho \nu \sigma }R^{\rho \sigma }-2R_{\mu
\sigma }R_{\phantom{\sigma}\nu }^{\sigma }+RR_{\mu \nu })-\frac{1}{2}%
L_{GB}g_{\mu \nu }  ~.
\end{equation}
By solving the Einstein equation (\ref{eqs}), the solution of the
TGBAdS black hole is given by~\cite{BD}
\begin{equation}\label{met}
  ds^2 =  - f(r) dt^2 +\frac{dr^2}{f(r)} + r^2 d\Omega_{3}^2.
\end{equation}
The metric function with two branches $\epsilon=\pm 1$ is given by
\begin{equation}
 f(r) =  k +\frac{r^2}{2c}\left[1+\epsilon\sqrt{1+\frac{4c}{3}\left(\frac{2\mu
 }{r^4}-\frac{3}{l^2}\right)}\right],
  \label{solf}
\end{equation}
where the signature $k$ classifies the horizon geometry depending
on $k= -1$ (hyperbolic), 0 (flat), 1
(spherical)~\cite{RCAI,CNO,Neupane}.
 Here $\mu$  is the mass parameter related to the black hole
mass $M=\mu{\cal A}_{3}/8\pi G_5$ where ${\cal A}_{3}=2\pi^2$ is
the area of unit three sphere. The GB black hole solution with
$k=1$ spherical horizon  was first found by Boulware and
Deser~\cite{BD}. Note that although our conventions follow
Ref.~\cite{Hira}, there exits a clear correspondence to
Ref.~\cite{DMMS2}: $c \to \tilde{\alpha}$,~$\mu \to 3m/2$.

On the other hand, the metric (\ref{met})  goes to AdS space
asymptotically. In the limit of $r \rightarrow \infty $, the
metric function takes the form  as
\begin{equation}
f_\infty(r) =
\frac{r^2}{2c}\left(1 + \epsilon \sqrt{1-\frac{4c}{l^2}}\right)\equiv
 \frac{r^2}{l^2_{eff}} \label{ads}
\end{equation}
with the effective AdS$_5$ curvature radius
\begin{equation}
l^2_{eff}= \frac{l^2}{2}\Bigg[1 - \epsilon \sqrt{1-\frac{4c}{l^2}}\Bigg].
\end{equation}
We note  that this  metric is well-defined on the boundary at the
infinity
 if
\begin{equation}
c \le l^2/4.
\label{cond}
\end{equation}
Therefore, one has to consider $c$ satisfying the above bound
for the $c>0$ case with $\epsilon=\pm 1$,
while the $\epsilon=-1$ solution is only possible
without any bound for the case of $c<0$.

\section{Thermodynamics of  GBAdS black holes}

From now on, we consider  the $k=1$ with $\epsilon=-1$, while the
$k=-1$ case will be treated in Appendix A. The zeros of $f(r)$
determine the locations of the horizons. In the five dimensions,
there is a single horizon at
\begin{equation}
r_h^2 (\mu,c) = \frac{l^2}{2}\left(-1 + \sqrt{1+\frac{4(2\mu
-3c)}{3l^2}}\right). \label{bhorizon}
\end{equation}
We note that in order to have a real solution $r_h \geq 0$,  $\mu
\geq \frac{3}{2}c $ is required for the case of $c>0$. Otherwise,
there is no event horizon for black hole. Now, let us derive the
mass parameter  as a function of the horizon radius $r_h$. From
Eq. (\ref{bhorizon}), we have the  mass parameter
\begin{equation}
\mu(r_h,c) = \frac{3}{2} \left( \frac{r_h^4}{l^2}+r_h^2 + c
\right). \label{mu}
\end{equation}
In order to obtain the structure of the GBAdS, we wish to find the
extremal black hole. From the conditions of $f=0$ and $f'=0$, we
find that the degenerate horizon is located at
\begin{equation}
r_e \equiv r_h = 0,
\end{equation}
which is confirmed by noting $r_h^2 \to 0$ as $\mu \to 3c/2$ in
Eq. (\ref{bhorizon}). In this case, the mass parameter is given by
\begin{equation}
\mu_e\equiv \mu(0,c)=\frac{3}{2}c.
\end{equation}
Here we point out  that $\mu_e$ is  a mass gab~\cite{Cho} as well
as  a mass of the  extremal GBAdS.
 Then, the bound of mass is given by
\begin{equation}
\mu_e \le \mu(r_h,c)\le \mu(r_h,\frac{l^2}{4}).
\end{equation}
 On the other hand, for the $c<0$ case, there is no
extremal black hole, but there is a bound for  mass
\begin{equation}
\mu \ge -\frac{3}{2}|c|.
\end{equation}
Here we have negative mass of $ \mu<0$ for $r_h<l_0$ with $l^2_0
\equiv (l^2/2)[-1+\sqrt{1-4c/l^2}]$, zero mass of $\mu=0$ at
$r_h=l_{0}$, and positive mass of $\mu>0$ for $r_h>l_0$ as
depicted in Fig. 1.
\begin{figure}[t!]
   \centering
   \includegraphics{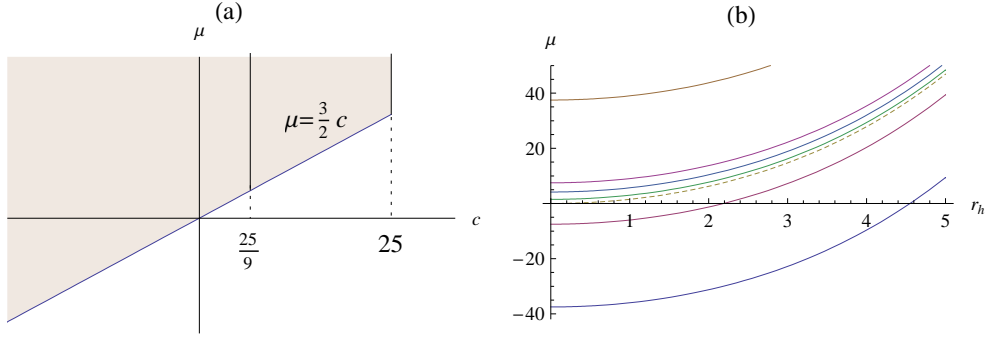}
\caption{(a) The mass function graph $\mu$ vs $c$ with $l=10$.
$\mu=3c/2$ is the lower bound for $c < 0$, while it corresponds
to the extremal black hole for $c>0$.
 The naked singularity regions (NS) are white areas below solid oblique of
$\mu=3c/2$ and  the forbidden region appears for $c>l^2/4$. (b)
The mass function graphs $\mu$ vs $r_h$  with $l=10$ from top to
bottom: $c=25$(upper bound), $5, 25/9, 1, 0$(SAdS),$ -5,
-25$. For the case of $c>0$, we have always positive mass and the
extremal black holes at $r_h=0$. For $c< 0$,
we have negative, zero, and positive masses.} \label{fig1}
\end{figure}

Now, we are ready to drive the thermodynamic quantities of GBAdS.
The Hawking temperature  defined by $T_H=f'(r_h)/4\pi$ takes the
form
\begin{equation}
\label{aas2} T_H(r_h,c) = \frac{r_h^3}{\pi l^2 (r^2_h + 2c)} \left(
1+ \frac{l^2}{2 r^2_h}  \right).
\end{equation}
Note that from the condition $T_H(r_h,c)\geq 0$, $c$ should
satisfy the condition $c>0$ or $r_h^2 \geq -2c$ for the case of
$c<0$. Then, using the Eqs. (\ref{mu}) and (\ref{aas2}), the heat
capacity defined by $C(r_h,c)=(dM/dT_{H})_c$ is obtained to be
\begin{equation}\label{aac}
C =\frac{3 {\cal A}_3}{4G_5} \Bigg[\frac{ r_h \left(2 r^2_h + l^2
\right) \left(r^2_h + 2 c \right)^2 } { r^2_h(2r^2_h -  l^2) + 2 c
\left(6 r^2_h + l^2\right)}\Bigg].
\end{equation}
The heat capacity is very important to test the thermodynamic
stability of the black hole: if $C>0$, the black hole is
thermodynamically stable, while for $C<0$, the corresponding black
hole is unstable. If $C>0$, the global stability is guaranteed by
the condition of a negative free energy $F<0$. On the other hand, the
Bekenstein-Hawking entropy derived from the first-law of
thermodynamics takes the form~\cite{CRS}
\begin{equation}
S_{BH} =\int \label{aas}
\frac{dM}{T}=\int^{r_h}_0\frac{dr_h}{T}\Big(\frac{dM}{dr_h}\Big)=\frac{{\cal
A}_3r_h^3}{4G_5}\Big(1 + \frac{6c}{r^2_h}\Big),
\end{equation}
which shows obviously that the area-law of the entropy does not
hold for the GBAdS. Furthermore, we have always a negative entropy
for the $c<0$ case. On the other hand, the on-shell free
energy defined by $F(r_h,c)= M -M_e-T_{H}S_{BH}$ is  given by
\begin{equation}
\label{aaf} F = - \frac{{\cal A}_3}{16 \pi G_5(r^2_h
+2c)}\Bigg[\frac{r_h^6}{l^2}-r_h^4 +3c\Big(\frac{6r_h^4}{l^2} +
r_h^2 - 2c\Big)\Bigg]-\frac{{\cal A}_3 }{16 \pi G_5}c.
\end{equation}
Hereafter, we consider two cases of $c < 0$ and $c>0$ separately
because these provide two distinct branches. Actually, we show
that ``$c$" plays the role of negative mass for $c<0$, whereas
``$c$" plays the role of pseudo-charge  for $c>0$. However, if one
considers $c$ as the pseudo-charge, the extremal mass $M_e$ should be
included as the ground state in defining the free energy of Eq.
(\ref{aaf}) in the canonical ensemble~\cite{CEJM} similar to the RNAdS case.
This picture is important to understand the nature of black holes obtained
from the GB term.

 \begin{figure}[t!]
   \centering
   \includegraphics{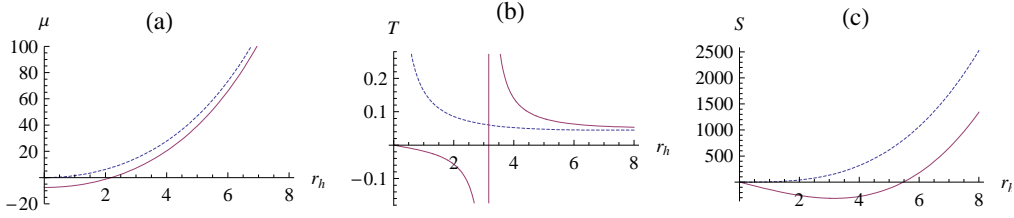}
\caption{For $c=-5$ GBAdS black hole, improper thermodynamic
quantities (the solid curves). (a) $\mu$ vs $r_h$ (b) $T$ vs
$r_h$, and (c) $S$ vs $r_h$. All show negative behaviors, which
are obstacles to define a proper black hole for $c<0$
thermodynamically. The dotted curves stand for the SAdS, showing
good features. } \label{fig2}
\end{figure}

 \begin{figure}[t!]
   \centering
   \includegraphics{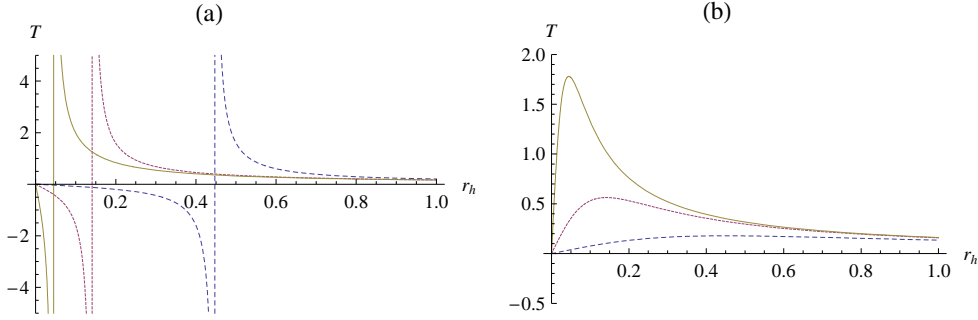}
\caption{The graphs of temperature vs horizon radius, showing that
we cannot obtain the SAdS in the limits of $c\to 0^\mp$. (a)
$c=-0.1$ for dashed, $c=-0.01$ for dotted, and $c=-0.001$ for
solid curves.  (b) $c=0.1$ for dashed, $c=0.01$ for dotted, and
$c=0.001$ for solid curves.} \label{fig3}
\end{figure}

\subsection{$c < 0$ GBAdS black holes: Not well-defined}
As is shown in Fig. 2, this $c < 0$ case shows a typical behavior of the
negative mass, negative temperature, and negative entropy
except the $r_h > l$ case of BBH. Explicitly, we have $\mu<0$ for
$r_h<l_0=(l/\sqrt{2})[-1+(1-4c/l^2)^{1/2}]^{1/2}$, $T<0$ for
$r_h<r_c=\sqrt{-2c}$ and $S<0$ for $r_h<\sqrt{-6c}$. These
negative values persist to any negative coupling.
This implies that the thermodynamics is not well-defined for the $c<0$
GBAdS. We call these the improper black hole because these can not
pass under the thermodynamic test.

The $c=0$ case is the SAdS whose temperature is given by
\begin{equation} \label{aasSt}
T^{SAdS}_{H}(r_h) = \frac{r_h}{\pi l^2}\left(1+ \frac{l^2}{2
r^2_h}\right).
\end{equation}
In this case, the entropy, heat capacity, and free energy  are
given by
\begin{eqnarray}\label{aaSbh}
S_{BH}^{SAdS}(r_h)&=&\frac{{\cal A}_3r_h^3}{4G_5}, \\
\label{aaSc}C^{SAdS}(r_h)&=& \frac{3 {\cal A}_3r_h^3}{4G_5}\Bigg[\frac{2r_h^2+l^2 }{2r_h^2 - l^2 }\Bigg], \\
\label{aaSf} F^{SAdS}(r_h)&=& -\frac{{\cal A}_3 r^2_{h}}{16 \pi
G_5}\Bigg[ \frac{r_{h}^2}{l^2}-1\Bigg].
\end{eqnarray}
For the SAdS case, we have a typical form of heat capacity,
showing the change from $-\infty$ to $\infty$ at the minimum
temperature point of  $ r_0 = l/\sqrt{2}$. Hence, this case has
two phases of negative ($-$) and positive $(+)$ heat capacities.

Let us check whether in the limit of  $c \rightarrow 0^-$, the
temperature $T_H$ could reduce to that of the SAdS. At the first
sight, the mathematical expression of Eq. (\ref{aas2}) superficially seems to
lead to Eq. (\ref{aasSt}). However, as depicted in Fig. 3-a, we
cannot arrive at the SAdS case as $c \to 0^-$ because the negative
temperature always appears unless $c=0$. The same also works for mass
$\mu$ and entropy $S_{BH}$. This feature persists in all $c<0$
black holes. As a result, these confirm thermodynamically that the SAdS black
hole could not be continuously reduced from the GBAdS.

\subsection{$ c> 0 $ GBAdS black holes}
For the $c>0$ case inspired by string theories, we have
completely different black holes.  Since the extremal black holes
are defined to have zero temperature, let us check that there is an
extremal solution at $r_h= r_e = 0$ with $\mu = \frac{3}{2}c$. In
this case, we extract out important points of the Davies' point (D)
~\cite{Dav,Pavon,JP} and minimum temperature point (0) from the
condition of $dT/dr_h=0$~\cite{Rmkp},
\begin{equation}
r^2_{D/0}=3\Bigg[\Big(\frac{l^2}{12}-c\Big) \mp
\sqrt{(c-l^2/4)(c-l^2/36)}\Bigg].
\end{equation}
These two points appear for the  $0 <c<l^2/36$ case, while the critical
(inflection) point appears at $r_h^2=l^2/6$ for $c=l^2/36$, and
any point does not exist for $l^2/36<c \le l^2/4$. Hence, we
expect that the BIAdS-like black hole is for $0
<c<l^2/36$~\cite{Myu}, the critical GBAdS is for $c=l^2/36$, and
the NBTZ-like black hole is for $l^2/36<c \le
l^2/4$~\cite{myungplb}. The global features of the Hawking
temperature depending on the parameter $c>0$ are shown in Fig. 4.
\begin{figure}[t!]
   \centering
   \includegraphics{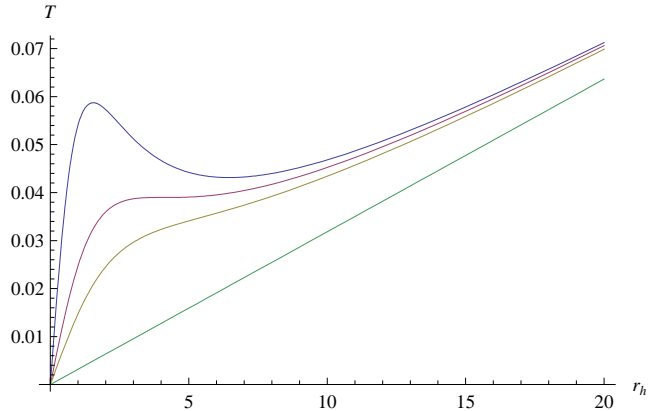}
\caption{Temperature graphs $T$ vs $r_h$  with $l=10$ for $c> 0$
from top to bottom: (a) $c=1$ ($bQ=0.5$ BIAdS) (b) $c=25/9$
(critical case), (c) $c=5$, (d) $c=25$ (NBTZ). All temperatures
are zero at the extremal point of $r_e=0$. } \label{fig4}
\end{figure}

In the case of $0 <c<l^2/36$, the two of Davies' point and local
minimum points appears. The graphs of the heat capacity depending
on the parameter $c>0$ are shown in Fig. 5. We find  that three
phases of $+-+$ for $0 <c<l^2/36$ with two blow-up points at
Davies' and minimum points (See Fig.5a). Here we define four  different black
holes~\cite{mkp4}: extremal black hole with zero heat capacity
(EBH$_0$), small black hole with positive heat capacity (SBH$_+$);
intermediate black hole with negative heat capacity (IBH$_-$);
large black hole with positive heat capacity (LBH$_+$). In fact,
these play the important roles in connection with the dynamical
stability. Furthermore, the $\delta$-function phase appears for
$c=l^2/36$ (See Fig.5b) and a single positive phase appears for $ l^2/36<c  \le
l^2/4$ (See Fig.5c and Fig.5d).

We note that for  $r_h > l$, we obtain  the big black hole
(BBH$_+$) as the asymptotic forms of the GBAdS and SAdS
\begin{equation} \label{Bbh}
T_{H}^{BBH}=\frac{r_h}{\pi l^2},~S_{BH}^{BBH}=\frac{ {\cal A}_3
r^3_h}{4G_5},~C^{BBH}=3S_{BH}^B,~ F^{BBH}=-\frac{ {\cal A}_3
r^4_h}{16 \pi G_5l^2}.
\end{equation}
These quantities are very useful for discussing the connection
between thermodynamic stability and QN modes. For this BBH$_+$,
the global thermodynamic stability is guaranteed because of
$C^{BBH}>0$ and $F^{BBH}<0$. However, for large black hole
(LBH$_+$), the global thermodynamic stability is not guaranteed
because for $r_0<r_h<r_1$ one has $C>0$ as shown in Fig. 5 and
$F>0$ as shown in Fig. 6. Here $r_1~(r_0)$ is the largest (next smaller) root
determined from the condition of $F=0$.
\begin{figure}[t!]
   \centering
   \includegraphics{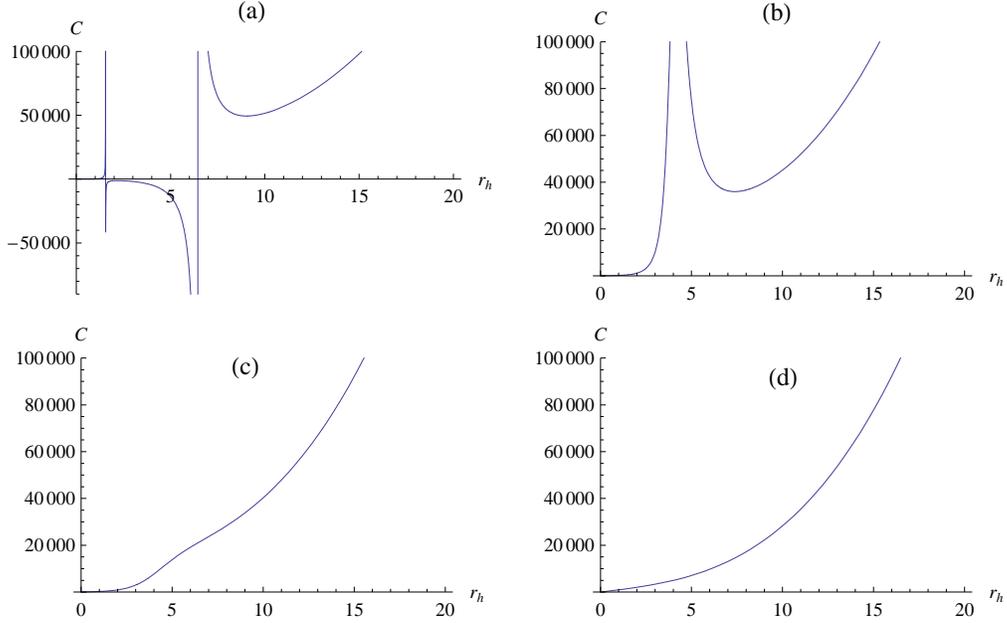}
\caption{Heat capacity graphs $C$ vs $r_h$ with $l=10$ for $c> 0$:
(a) $c=1$ ($bQ=0.5$ BIAdS) (b) $c=25/9$ (critical case), (c)
$c=5$, (d) $c=25$ (NBTZ).} \label{fig5}
\end{figure}
\begin{figure}[t!]
   \centering
   \includegraphics{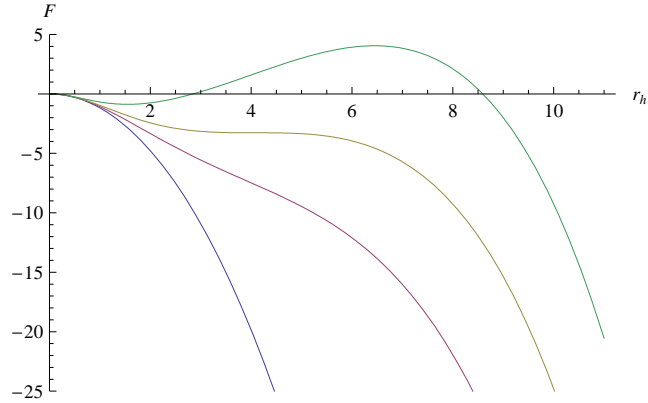}
\caption{Free energy graphs $F$ vs $r_h$ with $l=10$ for $c> 0$
from top to bottom: (a) $c=1$ ($bQ=0.5$ BIAdS) (b) $c=25/9$
(critical case), (c) $c=5$, (d) $c=25$ (NBTZ). } \label{fig6}
\end{figure}

\section{Thermodynamic stability and dynamical stability }
In this section we discuss the connection between thermodynamic
 and dynamical stability~\cite{HNO}.
 This is known as the correlated stability conjecture~\cite{GM,Reall}.
 First of all, the criterion of local thermodynamic
 stability is that $C>0~(C<0)$ denote stable (unstable) black
 holes. On the other hand, the dynamical stability is determined
 by investigating the time-independent $s$-mode perturbation
 around the GBAdS background. These perturbations are given by
 \begin{equation}
 h_a~^b={\rm diag}\Big[H_{tt}(r),H_{rr}(r),K(r),K(r),K(r)\Big].
\end{equation}
In the Einstein gravity, there exist two independent modes, which are trace and
traceless modes. The former is related to a conformal rescaling
and it has a ghost-like kinetic term, while the latter gives
negative modes. In order to compute the
eigenvalues $\lambda$ of the Lichnerowicz operator $\Delta$, one
should solve the tensor eigenvalue equation
\begin{equation}
 \Delta_{~b~d}^{a~c}~  h_c~^d=\lambda ~ h^a_{~b}
\end{equation}
with proper boundary conditions at infinity and horizon. Recently,
Hirayama has studied the asymptotic behaviors of the $s$-mode perturbation
at the horizon and the infinity in $r$,
and solved this equation numerically~\cite{Hira}. Hence,
we simply use this data to discuss the connection between dynamical
and  thermal stability. In this case, the criterion of dynamical
stability means  that $\lambda>0~(\lambda<0)$ denote stable
(unstable) black holes. In this section, we separate the whole
region of $c$ into the following three cases because each case has different
phase for the heat capacity.

\subsection{$c< 0$ GBAdS black hole}
In this case, we may have a direct connection between
thermodynamic and dynamic stability~\cite{Hira}.  We
have the SBH$_-$ for $r_c<r_h<r_0$, while the LBH$_+$ appears for
$r_h>r_0$. On the dynamical perturbation side, we may have the
same result, which shows the unstable black hole with $\lambda<0$
for $r_c<r_h<r_0$ and the  stable black hole with $\lambda>0$ for
$r_h>r_0$. However, the connections  of $C > 0 \leftrightarrow
\lambda> 0$ and $C < 0 \leftrightarrow \lambda< 0$ are meaningless
because this black hole is not well-defined thermodynamically.
This clearly modifies  the result of Ref.~\cite{Hira}, which stated
that the correlated stability conjecture holds for the $c< 0$
GBAdS black hole.

\subsection{$0<c< l^2/36$  GBAdS black holes}
The $0<c< l^2/36$ case seems to not provide a direct connection
between thermodynamic and dynamical stability~\cite{Hira} because
this case has four different phases of the heat capacity~\cite{mkp4}.
These are EBH$_0$,  SBH$_+$, IBH$_-$, and LBH$_+$. Up to now, we
regard the EBH$_0$,  extremal black hole with zero heat capacity
as the stable remnant of black hole. According to the perturbation
calculation, the corresponding eigenvalue is a positive constant,
which confirms that the EBH$_0$ is thermodynamically stable even
it has zero heat capacity.  For the LBH$_+$, there exists a direct
connection of $C>0\leftrightarrow\lambda>0$. However, there is no
such relation for the SBH$_+$ and IBH$_-$ cases.
Thermodynamically, the SBH$_+$ is quite different from SBH$_-$.
The former appears as the near-extremal black hole in the RNAdS, while
the latter is  the small unstable black hole, mediating the
Hawking-Page transition (HP2) in the SAdS. On the other hand, the IBH$_-$
appears the intermediate unstable black hole in the RNAdS.

Then, one may conjecture that two black holes of the SBH$_+$ and IBH$_-$
are closely related to the unexplored transition at the Davies' point ($r_h=r_D$).
Although the SBH$_+$ is thermodynamically stable, while
the IBH$_-$ is unstable as the index indicates, there
is no single value for $\lambda$ for these cases on the perturbation side.
That is, one has multiple values of $\{\lambda_i\}$ for a given horizon radius
$r_h$ around the Davies' point. Hence we do not confirm the
connection between thermodynamic and dynamical stability
for the SBH$_+$ and IBH$_-$ cases. This means that the Davies'
point is quite different from the minimum temperature point
($r_h=r_0$) even they give rise to the same blow-up of the heat
capacity in the GBAdS~\cite{mkp4} and RNAdS~\cite{Myu}. Hence, we
do not make any decisive connection between thermodynamic and
dynamical stability for the $0<c< l^2/36$  GBAdS. Therefore,
the correlated stability conjecture does not work for this black hole,
as it does not work for the RNAdS.

\subsection{$l^2/36<c\leq l^2/4$ GBAdS black holes}
We have a single phase only for the $l^2/36<c\leq l^2/4$ GBAdS. This
case is simple and similar to that of the NBTZ without SBH$_-$.
Although the SBH$_-$ plays a role of the mediator in the HP2,
it is absent here. The connection is clearly defined:
thermodynamically stable black hole
($C>0$) $\longleftrightarrow$ stable black hole ($\lambda$=positive
constant)~\cite{Hira}. The correlated stability conjecture does
work for this black hole well.

\section{Thermodynamic stability and QN modes }
First, let us discuss the thermodynamic fluctuations of black
holes. There exist already several works, which show that for a
large black hole in AdS spacetimes, the Bekenstein-Hawking entropy
receives logarithmic corrections due to thermodynamic
fluctuations~\cite{KM1,KM2,Car,GKS,BS,BanerjeeM}. The suggested
formula takes the form
\begin{equation}
 \label{CEN}
S=S_{BH}-\frac{1}{2} \ln [C] + \cdots,
\end{equation} where $C$ is the
 heat  capacity of Eq. (\ref{aac}) and $S_{BH}$
denotes the uncorrected Bekenstein-Hawking entropy in Eq.
(\ref{aas}). Here, an important point is that in order for Eq.
(\ref{CEN}) to make sense, $C$ should be positive. Hence, we can
not make any correction to  the entropy  for the EBH$_0$ and
IBH$_-$, while we could make  correction to the SBH$_+$ and LBH$_+$.
As a limiting case, the 5D Schwarzschild black hole, which is
asymptotically flat, has a negative heat capacity of
$C^{Sch}=-3S^{Sch}_{BH}$~\cite{DMB}. This means that the
Schwarzschild black hole is never in thermal equilibrium and it is
always unstable against thermal fluctuations. Thus, we do not make
any correction to the entropy of the Schwarzschild black hole. On
the other hand, we have $C>0$ for the BTZ black hole, which shows
that the entropy correction is  always possible to occur. This
means that the thermodynamic fluctuations on the black hole with
the positive heat capacity leads to the logarithmic correction to the
Bekenstein-Hawking entropy.

On the other hand, it is well known that if one perturbs a black
hole, the surrounding geometry will ring (undergo damped
oscillations)~\cite{Kokk}. These damped oscillations known as
``quasinormal(QN) modes" are entirely fixed by the thermodynamic
quantities of black hole and are independent of the initial
perturbations. In general, their QN frequencies are complex and discrete,
\begin{equation}
\omega=\omega_R-i \omega_I,
\end{equation} which describe the decay
of external perturbations outside the event horizon. However, as
far as we know,  their analytic expressions are known for two
limited cases: NBTZ in three dimensions~\cite{BSS} and topological
massless black hole (TMBH) in higher dimensions~\cite{BM}. For the
big black holes in AdS spaces, the QN frequencies are known to
take analytic form approximately~\cite{HH,Siop}. However, for
small and intermediate black holes, there exists no analytic form.
This suggests  a close connection between the thermodynamic
stability and QN modes.

It seems appropriate to mention  the connection between global
thermodynamic stability and QN modes for BBHs in AdS spacetimes.
If $C>0$, the global thermodynamic stability (GTS) is guaranteed
by the condition of a negative free energy $F<0$. Actually, the
BBH$_+$ with $r_h >l$ satisfies the GTS. As shown in Eq. (\ref{Bbh}),
this case always has negative free energy. Hence, we
conjecture the important connection between GTS and QN modes:

 {\it If a black hole is big and
thus globally stable, its QN modes may take analytic expressions}.

The typical examples are 3D NBTZ and 5D TMBH for completely analytic
expressions and the big SAdS for approximately analytic
expressions. The first two expressions for scalar perturbations
are given by~\cite{BSS,BM,Siop,ML}
\begin{eqnarray} \label{qnmn}
\omega^{NBTZ}_s&=& 4\pi  T_H^{NBTZ}\Big[\pm \ell- i
(n+ 1)\Big],\\
\label{qnmt} \omega^{TMBH}_s&=&2 \pi T_H^{TMBH}
\Big[\pm\xi_s^{TMBH}-  2 i \Big(n+ \frac{1}{2}\Big)\Big],
\end{eqnarray}
where $T_H^{NBTZ}=\frac{r_h}{2\pi l^2}$, $\ell$ with $\nabla^2
Y=-\ell^2Y$ on $S^1$ and $T_H^{NBTZ}=\frac{1}{2\pi l}$,
$\xi_s^{TMBH}=\sqrt{k_s^2-1/4}$ with $\nabla^2 Y=-k_s^2Y$ on
$S^3$. For the NBTZ, we have a massless scalar, while we consider
the massive scalar with $m^2l^2=-4$ for the TMBH. The QN
frequencies for BBH$_+$  takes the following form for a massless
scalar~\cite{MSIO}:
\begin{equation} \label{qnmb}
\omega^{BBH}_s \simeq 8 \pi T^{BBH}_H \Big[ \pm n-i n\Big],
\end{equation}
where $T_H^{BBH}=\frac{r_h}{\pi l^2}$.

Parallel to the previous section, we also separate the whole
region of $c$ into the following three cases.

\subsection{$c< 0$ GBAdS black holes}
In this case, we may have a connection between thermodynamic
stability and QN modes for the BBH$_+$. We assume that QN frequencies
take approximately analytic form. However, as emphasized
before, this black hole is not well-defined thermodynamically.
Hence, it is meaningless to consider this connection for the $c<
0$ GBAdS black hole.

\subsection{$0<c\leq l^2/36$ GBAdS black holes}

For the SBH$_+$ and IBH$_-$ cases, we can not expect to have
analytic form of QN frequencies because their connection between
thermodynamic and dynamical stabilities is not yet  established.
However, for the BBH$_+$, we may have approximately analytic form
of QN frequencies like Eq. (\ref{qnmb}).  Finally, we expect that
the QN frequencies are changed drastically around the Davies'
point for this case because the same thing happens for the
RNAdS~\cite{JP,BC}.

\subsection{$l^2/36<c\leq l^2/4$  GBAdS black holes}
We expect that its QN modes take analytic form because this case has a
single phase of the positive heat capacity. Specifically, for the upper bound
case of $c=l^2/4$, we strongly expect to have the presence of
analytic expressions like Eq. (\ref{qnmn}) because this case
corresponds to the NBTZ~\cite{BSS}.

\section{Summary and discussions}
We have considered the GBAdS black hole in five dimensions to
study its thermodynamics thoroughly. First of all, we have pointed
out that this black hole does satisfy the first-law of
thermodynamics, but its entropy does not satisfy the area-law. This
is mainly because we have used the first-law to derive the
entropy.

We could not find a thermodynamically well-defined black hole for
the $c<0$ case where $c$ may play the role of negative mass. Although this
seems to be the SAdS-type black hole, all of its mass,
temperature, and entropy are negative unless $c=0$. Moreover,
the role of negative mass ``$c$" becomes clear when introducing the $k=-1$
topological GBAdS (TGBAdS) in Appendix A.

On the other hand, for the $c>0$ case, the
$bQ=0.5$ BIAdS will be used to study the $0<c<l^2/36$ GBAdS
because their thermodynamic properties are the nearly same. This
corresponds to the black hole inspired by the string theories. In
this case, $c$ plays the role of a pseudo-charge. The role of
charge ``$c$" becomes clear when considering the GBRNAdS black
hole in Appendix B. We clarify the role of ``$c$" in Table 1.

\begin{table}
\caption{The role of the Gauss-Bonnet coupling constant $c$ for
various black holes.}
\centering
\begin{tabular}{|c|c|c|c|}
  \hline
  black holes & $c<0$ & $c>0$ & reference \\
  \hline
  GBAdS & negative mass & pseudo-charge & SAdS/ BIAdS \\
  TGBAdS& negative mass & $\cdot$ &  TAdS/~~~~~$\cdot$~~~~~\\
  GBRNAdS&   $\cdot$ & charge& ~~~$\cdot$~~~~/~RNAdS \\
  \hline
\end{tabular}
\end{table}

Furthermore, we could not obtain the SAdS in the limits of $c\to
0^{\mp}$. This contradicts to the view that the GBAdS is
the deformed SAdS. The origin of this problem is the order of
taking the $c\to0$ limit in calculation of thermodynamic
quantities. If this limit is taken before the calculation of
thermodynamic quantities, then we find thermodynamic
quantities of the SAdS very well. Actually, this corresponds to turning off
the GB term $L_{GB}$ in the action. On the other hand, if this
limit is taken after the calculation of thermodynamic quantities,
then we could not find the thermodynamic quantities of the SAdS. This
implies that the  order of taking the $c\to0$ limit is very
important to recover the SAdS from the GBAdS.

In addition, we have clarified the connection between thermodynamic and
dynamical stability, known as the correlated stability conjecture.
We have found that this conjecture is valid for the $l^2/36 <c \le
l^2/4$ GBAdS only because this case has a single positive phase of
the heat capacity. On the other hand, for the $0 <c \le l^2/36$ GBAdS,
this conjecture does not work because this has four different phases of
the heat capacity, showing that it is not easy to make a direct connection between
thermodynamic and dynamical stability.

Finally, we have newly proposed the important connection between the
global thermodynamic stability (GTS: $C>0, F<0$) and the analytic
expressions of QN frequencies: {\it If the black hole satisfies
the GTS, its QN modes may have the analytic form}. The conjecture
works for the black hole in asymptotically AdS spacetimes.

\renewcommand{\theequation}{A\arabic{equation}}
\setcounter{equation}{0}  
\section*{Appendix A: Thermodynamic quantities of $k=-1$ TGBAdS black holes}

\begin{figure}[t!]
   \centering
   \includegraphics{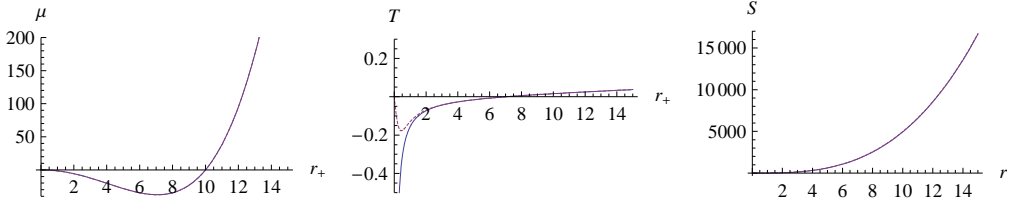}
\caption{Thermodynamic quantities of mass $\mu$, temperature $T$,
and entropy $S$ for $k=-1$ TGBAdS black hole. The dotted curves
represent the $k=-1$ TAdS black hole with $c=0$, while  the solid
curves denote $k=-1$ TGBAdS with $c=-0.10$. } \label{fig7}
\end{figure}

In this case with $c<0$,  we have the relevant thermodynamic quantities
\begin{equation}
\mu(r_h)= \frac{3 }{2} \left(\frac{r^4_h}{l^2}-r^2_h+c\right),
\end{equation}

\begin{equation}
T(r_h)=\frac{r^3_h}{\pi
l^2\left(r^2_h-2c\right)}\left(1-\frac{l^2}{2r^2_h}\right),
\end{equation}

\begin{equation}
S(r_h)=\frac{{\cal A}_3 r^3_h
}{4G_5}\left(1-\frac{6c}{r^2_h}\right),
\end{equation}
The important observation is that there is no blow-up point of $T$ in Eq. (A2), and
the entropy is always positive as is shown in Eq. (A3). These improve the troubles related to the
blow-up temperature and the negative entropy in the $c<0$ GBAdS
black hole. These are depicted in Fig. 7.

\renewcommand{\theequation}{B\arabic{equation}}
\setcounter{equation}{0}  
\section*{Appendix B: Thermodynamic quantities of GBRNAdS black holes }
In this case with $c>0$, the metric function takes the form
\begin{equation}
f(r)=1+\frac{r^2}{2c}\left(1-\sqrt{1+\frac{2c}{3}\left(\frac{4\mu
}{r^4}-\frac{6}{l^2}\right)-\frac{9q^2}{2r^6}}\right),
\end{equation}
which is obtained by adding  $-\frac{1}{16\pi G_5}F^2$  to the
action (\ref{baction}). Then, the relevant thermodynamic
quantities of mass $\mu$, temperature $T$, and the entropy $S$
are given by \cite{RCAI}

\begin{equation}
\mu(r_h,c,q)= \frac{3}{2} \left(\frac{r^4_h}{l^2}+
r^2_h+\frac{9q^2}{8c}\frac{1}{ r^2_h}\right),
\end{equation}
\begin{equation}
T(r_h,c,q)=\frac{r^3_h}{\pi l^2 \left(r^2_h+2c\right)}
               \left(1+\frac{l^2}{2r^2_h}-\frac{9q^2}{16c}\frac{l^2}{
               r^6_h}\right),
\end{equation}
\begin{equation}
S(r_h,c,q)=\frac{\pi^2}{2} r^3_h \left(1+\frac{6c}{r^2_h}\right).
\end{equation}
From Fig. 8, we observe the role of $c$ as the square of charge
$q^2$  because it appears as a combination of $q^2/c$ in Eqs. (B2)
and (B3).

\begin{figure}[t!]
   \centering
   \includegraphics{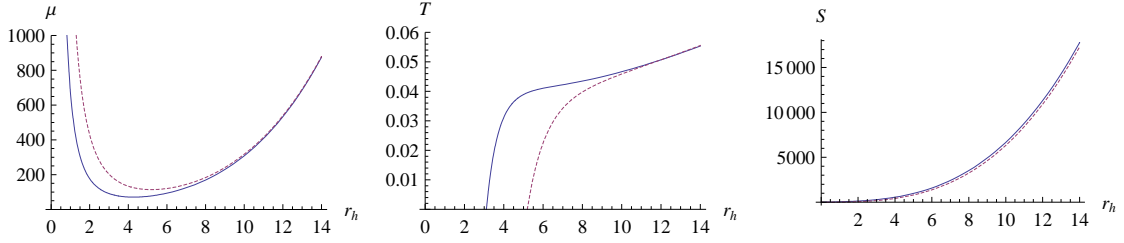}
\caption{Thermodynamic quantities of mass $\mu$, temperature $T$,
and entropy $S$ for GBRNAdS black hole. The dotted curves
represent the RNAdS black hole with $c=0$, while  the solid curves
denote GBRNAdS with $c=0.1$. Two shows the nearly same behavior.}
\label{fig8}
\end{figure}

\medskip

\section*{Acknowledgments}
Y. S. Myung  was supported by the Korea Research Foundation (KRF-2006-311-C00249)
funded by the Korea Government (MOEHRD). Y.-W. Kim was supported by the Korea
Research Foundation Grant funded by Korea Government (MOEHRD):
KRF-2007-359-C00007. Y.-J. Park was supported by
the Korea Science and Engineering Foundation (KOSEF) grant
funded by the Korea government (MOST) (R01-2007-000-20062-0).

%
%

\end{document}